\documentclass[twocolumn]{aastex62}
\usepackage{apjfonts}
\usepackage{amsmath}

\newcommand{\Hi}{\hbox{H{\sc i}}}

\begin{document}

\title{Local star-forming galaxies build up central mass concentration most actively near $M_{\ast}=10^{10}M_{\sun}$}
\author{Zhizheng Pan}
\email{panzz@pmo.ac.cn}
\affiliation{Purple Mountain Observatory, Chinese Academy of Sciences, 10 Yuan Hua Road, Nanjing, Jiangsu 210033, China}
\affiliation{School of Astronomy and Space Sciences, University of Science and Technology of China, Hefei, 230026, China
}

\author{Xianzhong Zheng}
\affiliation{Purple Mountain Observatory, Chinese Academy of Sciences, 10 Yuan Hua Road, Nanjing, Jiangsu 210033, China}
\affiliation{School of Astronomy and Space Sciences, University of Science and Technology of China, Hefei, 230026, China
}

\author{Xu Kong}
\affiliation{School of Astronomy and Space Sciences, University of Science and Technology of China, Hefei, 230026, China
}
\affiliation{CAS Key Laboratory for Research in Galaxies and Cosmology, Department of Astronomy, \\
University of Science and Technology of China, Hefei, Anhui 230026, China}

\begin{abstract}
To understand in what mass regime star-forming galaxies (SFGs) build up central mass concentration most actively, we present a study on the luminosity-weighted stellar age radial gradient ($\nabla_{\rm age}$) distribution of $\sim3600$ low-redshift SFGs using the MaNGA Pipe3D data available in the SDSS DR17. The mean age gradient is negative, with $\nabla_{\rm age}=-0.14$~log Gyr/$R_{\rm e}$, consistent with the inside-out disk formation scenario. Specifically, SFGs with positive $\nabla_{\rm age}$ consist of $\sim 28\%$ at log$(M_{\ast}/M_{\sun})<9.5$, while this fraction rises up to its peak ($\sim 40\%$) near log$(M_{\ast}/M_{\sun})=10$ and then decreases to $\sim 15\%$ at log$(M_{\ast}/M_{\sun})=11$. At fixed $M_{\ast}$, SFGs with positive $\nabla_{\rm age}$ typically have more compact sizes and more centrally concentrated star formation than their counterparts, indicative of recent central mass build-up events. These results suggest that the build-up of central stellar mass concentration in local SFGs is mostly active near $M_{\ast}=10^{10}M_{\sun}$. Our findings provide new insights on the origin of morphological differences between low-mass and high-mass SFGs.
\end{abstract}


\section{Introduction} \label{sec:intro}
The origin of galaxy morphology is a fundamental puzzle of galaxy formation and evolution. Bulges, characterized by dense spheroidal swarm of stars, are commonly found in the centers of galaxies. The origin of bulge stars could be quit complicated. Various physical processes such as mergers \citep{Hopkins 2009,Hopkins 2010}, giant clump migration \citep{Dekel 2022}, disk instability induced central star formation \citep{Kormendy 2004,Athan 2005} and rapid early-on star formation \citep{Okamoto 2013} are considered to play a role in bulge formation, although their relative contributions are still unclear \citep{Tacchella 2019,Huko 2023,Boecker 2023}.

In the past two decades, much efforts are made to understand the physics behind the empirical relationship linking the star formation histories of galaxies and their morphologies. Observational studies show that star formation quenching is associated with the presence of a bulge component, while galaxies with bulges are not necessarily to be quenched \citep{Bell 2012, Fang 2013,Barro 2017}. Multiple lines of evidence suggest that a dissipative core-building event ("compaction") that feeds an active galactic nucleus (AGN) can quench star formation \citep[e.g.,][]{Bluck 2014,Wang 2018,Woo 2019,Xu 2021}. Although the relation between star formation quenching and the presence of a bulge is still a matter of debate, the coupling of the two phenomena implies that the build-up of central mass concentration is a critical step in galaxy evolution.

In this work, we aim to explore in what mass regime SFGs build up central mass concentration most actively. To answer this question, one needs to investigate the spatially resolved star formation properties of a statistically large galaxy sample. MaNGA is the largest integral filed unit (IFU) spectroscopic survey to date, providing an excellent dataset suited for such studies. In this work, we present a study on the luminosity-weighted stellar age radial gradient ($\nabla_{\rm age}$) distribution of $\sim3600$ local SFGs ($0.01<z<0.15$) using data from the MaNGA survey available in the SDSS DR17 release. Throughout this paper, we adopt a concordance $\Lambda$CDM cosmology with $\Omega_{\rm m}=0.3$, $\Omega_{\rm \Lambda}=0.7$, $H_{\rm 0}=70$ $\rm km~s^{-1}$ Mpc$^{-1}$ and a \citet{Chabrier 2003} initial mass function (IMF).

\section{data}
MaNGA is an IFU spectroscopic survey for obtaining two-dimensional spectral mapping of $\sim 10,000$ nearby galaxies \citep{Bundy 2015}. The sizes of the IFUs vary from 19 to 127 fibers, and the effective spatial resolution is $2.^{\arcsec}5$ \citep{Law 2015}. The data products of the MaNGA project have been released in SDSS DR17.

We use the reduced data from the MaNGA Pipe3D value-added catalog \citep{Sanchez 2022}. The stellar mass ($M_{\ast}$) and photometric measurements are drawn from the NASA-Sloan Atlas catalog\footnote{nsatlas.org}, including SDSS $r-$band effective radius $R_{\rm e}$, S\'{e}rsic index $n$, minor-to-major axis ratio $b/a$. Star formation rate (SFR) is derived from the dust-corrected H$\alpha$ luminosity. This SFR is an upper limit to the real one, since all H$\alpha$ flux is integrated irrespective of the nature of ionization \citep{CanoDiaz 2016}. To remove edge-on objects, galaxies with $b/a>0.4$ are selected. We further restrict our analysis to galaxies with redshift $0.01<z<0.15$, $10^{9.0}M_{\sun}<M_{\ast}<M^{12}M_{\sun}$, and $\rm 10^{-5}M_{\sun}yr^{-1}<SFR<10^{3}M_{\sun}yr^{-1}$.  We also use the quality control flag QCFLAG to select datacubes that have good-quality analysis by the Pipe3D pipeline (with $\rm QCFLAG=0$). Galaxies observed with the 19-fiber IFUs are excluded from analysis because the low spatial-resolution IFUs may result in large uncertainties in the radial gradient measurements. The final sample contains 6916 galaxies.

The Pipe3D pipeline uses a library which compromises 273 simple stellar population (SSP) spectra, sampling 39 ages from 1 Myr to 13.5 Gyr and seven metallicities ($Z/Z_{\sun}$=0.006, 0.029, 0.118, 0.471, 1, 1.764 and 2.353), to model the observed spectra. This library is a subset of the MaStar SSP library, which is generated from an updated set of Bruzual and Charlot (BC03) stellar population synthesis model \citep{BC03,Plat 2019}. Pipe3D derived the luminosity- and mass-weighted parameters of the stellar population using the coefficients of stellar decomposition, with the equation:
\begin{equation}
\begin{split}
{\rm log} P_{\rm LW}=\sum\nolimits_{\rm ssp}w_{\rm ssp,L}{\rm log} P_{\rm ssp} \\
~{\rm log}P_{\rm MW}=\frac{\sum_{\rm ssp}w_{\rm ssp,L}\Upsilon_{\rm ssp,\bigstar}{\rm log} P_{\rm ssp}}{\sum_{\rm ssp}w_{\rm ssp,L}\Upsilon_{\rm ssp,\bigstar}},
\end{split}
\end{equation}

where $P_{\rm ssp}$ is the value of a particular parameter for each SSP, $w_{\rm ssp,L}$ is the weight contributed by each SSP, and $\Upsilon_{\rm ssp,\bigstar}$ is the mass-to-light ratio. The weight $w_{\rm ssp,L}$ is determined at a fixed spectral range of $5450<\lambda<5550$ \AA, similar to the $V$-band central wavelength. In the spectral modeling, the dust extinction law of \cite{Cardelli 1989} is adopted.

In Pipe3D, a radial gradient of a quantity of interest is computed as the slope of the linear fit of the quantity versus the galactic centric distance of the spaxel in unit of $R_{\rm e}$. First, the position angle, ellipticity, and $R_{\rm e}$ provided by the NSA catalog for each galaxy are used to create elliptical apertures of 0.15 $R_{\rm e}$ width, covering the galactic centric distance from 0 to 3.5 $R_{\rm e}$. The average value for each parameter at each distance is then estimated. For the radial distribution of each parameter, the slope of the average gradient is derived based on a linear regression of the considered parameter along the radius. The fitting is restricted between $0.5-2.0R_{\rm e}$. When the field of view (FOV) dose not reach 2.0$R_{\rm e}$, the regression is restricted to the largest distance covered by the FOV. In this work, we focus our analysis on the radial gradient of luminosity-weighted stellar age, $\nabla_{\rm age}$. The typical uncertainty of $\nabla_{\rm age}$ is around $\rm {0.05~log Gyr}/R_{e}$. More details of the MaNGA Pipe3D dataset please refer to \cite{Sanchez 2022}.

The morphological classification catalog of the MaNGA galaxies released by \cite{DS22} (DS22) is also used. This catalog provides a T-Type parameter (ranging from $-4$ to 9) for each MaNGA galaxy, which is obtained by training deep-leaning models based on the T-Types from \cite{NA10}. In general, T-Type<0 corresponds to early-type galaxies, while T-Type $\geq$ 0 correspond to late-type galaxies. For late-type galaxies, galaxies with 0$\leq$T-Type$\leq$3 tend to have larger bulge-to-disk ratios than those of T-Type > 3 (see Figure 22 of \citet{Fischer 2019}).

\begin{figure*}
\centering
\includegraphics[width=160mm,angle=0]{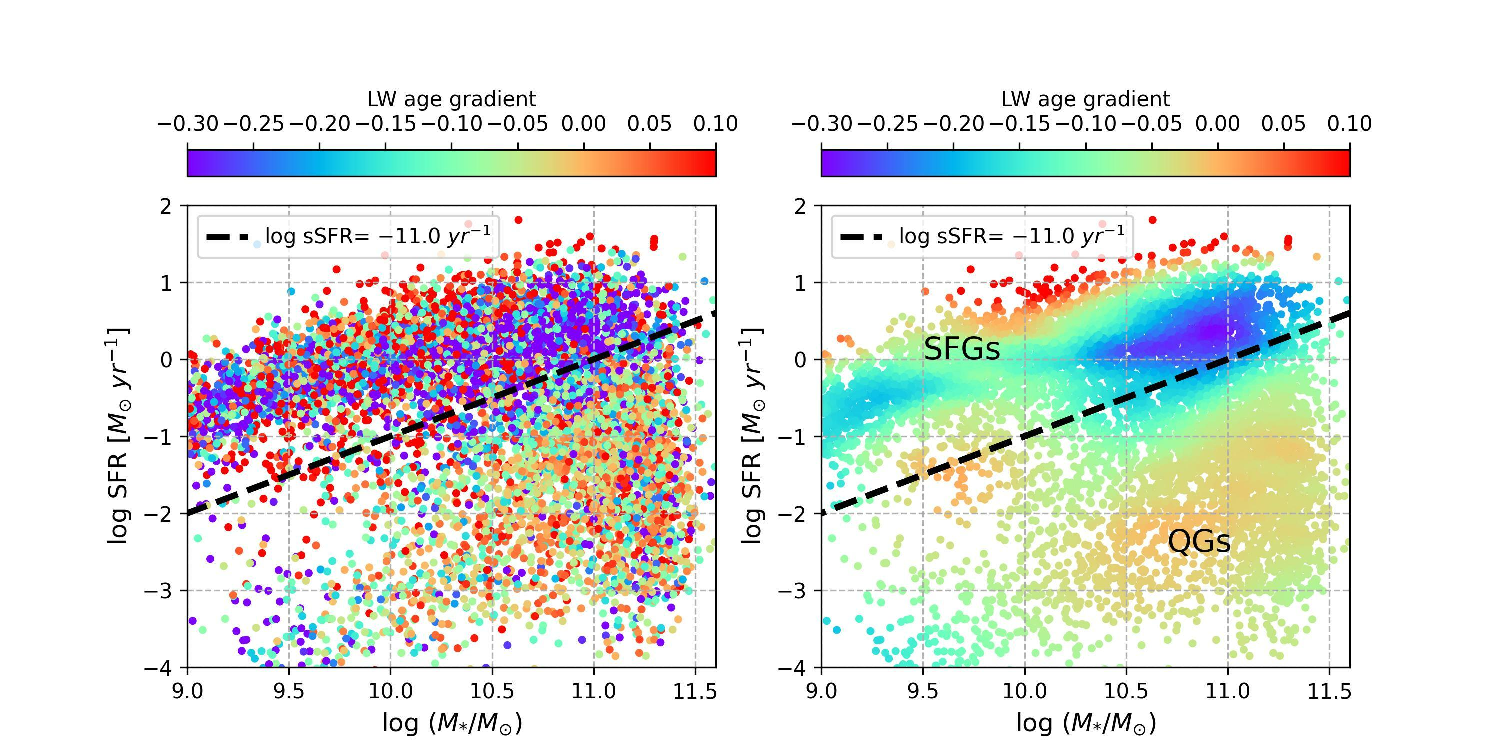}
\caption{left: the log${\rm SFR}-\rm log M_{*}$ plane. Symbols are color coded by the luminosity-weighted stellar age gradient. The dashed line indicates log sSFR$=10^{-11}yr^{-1}$, and SFGs are selected as those with SFRs above the line. Right: a LOESS smoothed version of the left panel.  }\label{fig1}
\end{figure*}

\section{Results}
\subsection{the distribution of luminosity-weighted age gradient of SFGs}
The luminosity-weighted stellar age radial gradient $\nabla_{\rm age}$ describes the spatial distribution of stellar age within a galaxy. In this work, we consider SFGs with $\nabla_{\rm age}>0$ (younger central stellar age than the outer region) are undergoing/or recently underwent central mass build-up events. This method should be valid for detecting central mass build-up events involving recent central star formation. As mentioned above, bulge stars could also be originated from stellar radial migration or \emph{ex situ} star formation (accreted from other galaxies). Such central mass build-up events are out of the scope of this work.

In the left panel of Figure~\ref{fig1}, we show the SFR$-M_{\ast}$ plane for the 6916 galaxies selected in Section 2.  Data points are color coded by $\nabla_{\rm age}$. In the SFR$-M_{\ast}$ plane, SFGs distribute along the star-formation main sequence (SFMS). To explore the $\nabla_{\rm age}$ distribution across the SFR$-M_{\ast}$ plane, we smooth the data using locally weighted regression method LOESS developed by \cite {Cleveland 1988}, with the \verb"PYTHON" code released by \cite{Cappellari 2013}. LOESS is developed to uncover underlying mean trends by reducing observational errors and intrinsic scatters. We adopt a smoothing factor of $frac=0.1$, and a linear local approximation.

The right panel of Figure~\ref{fig1} shows the LOESS-smoothed version of the SFR$-M_{\ast}$ plane. As can be seen, at fixed $M_{\ast}$, galaxies with the highest SFRs tend to have positive $\nabla_{\rm age}$. Such galaxies are likely undergoing the compaction processes \citep{Tacchella 2016,Ellison 2018}, thus having enhanced SFRs compared to the SFMS ridge line. Along the ridge line of the SFMS, galaxies tend to have negative $\nabla_{\rm age}$. At log$(M_{\ast}/M_{\sun})>10.5$, SFGs have steep and negative $\nabla_{\rm age}$. This is consistent with the observation that massive SFGs typically harbor old bulge components \citep{Perez 2013,Pan 2015,Pan 2016, Belfiore 2017}. Interestingly, SFGs near log$(M_{\ast}/M_{\sun})=10$ show shallower $\nabla_{\rm age}$ compared to those in the low-mass and high-mass regimes. In the low-SFR regime, galaxies generally have very shallow $\nabla_{\rm age}$. This is because the low-SFR regime is dominated by early-type galaxies, which typically exhibit very shallow age gradients \citep{Parikh 2021}.

\begin{figure*}
\centering
\includegraphics[width=160mm,angle=0]{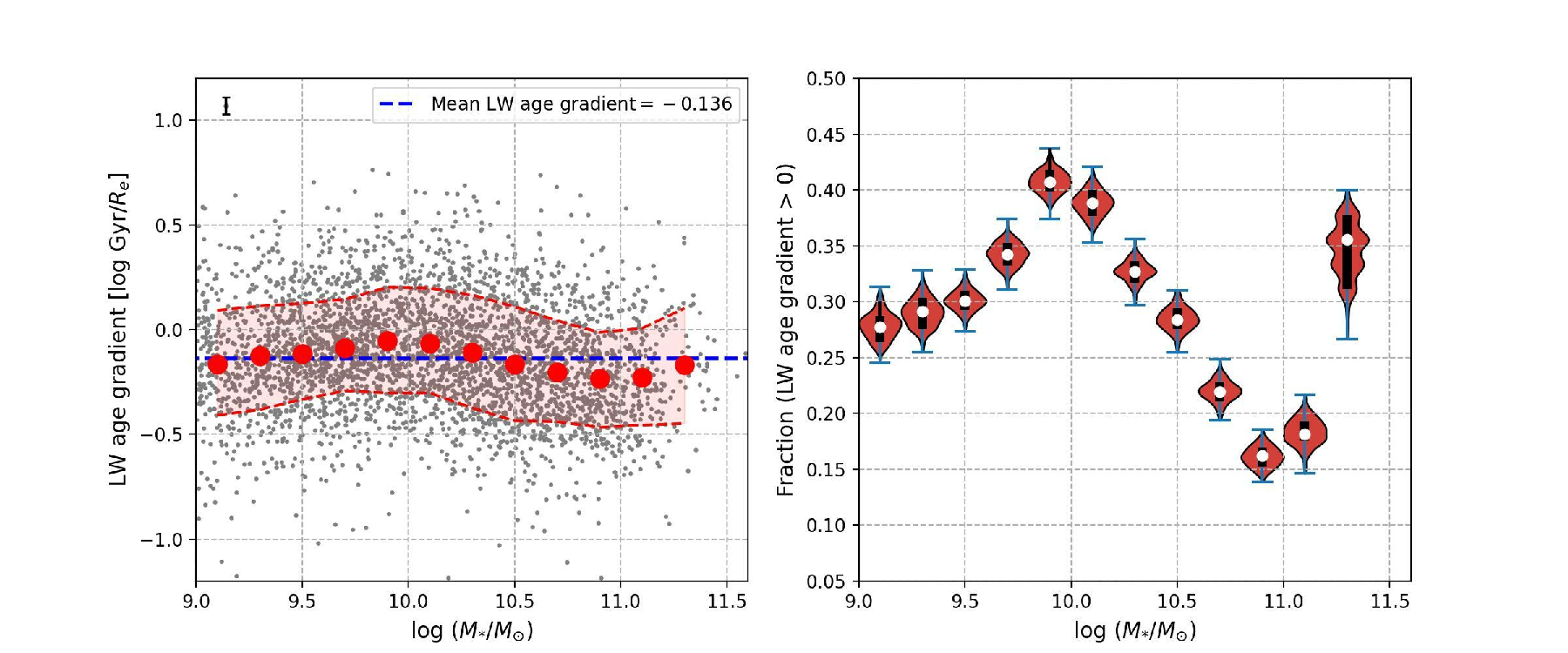}
\caption{Left: The luminosity weighted age gradient as a function of $M_{\ast}$ for SFGs. The mean age gradient is marked by the blue dashed line. Large symbols show the median $\nabla_{\rm age}$ values of each stellar mass bin, with a bin size of $\Delta M_{\ast}=0.2~$dex. The shaded region indicates the $16-84$ percentile region.  Right: violin plot showing the distribution of $F_{\rm positive}$ at each stellar mass bin.}\label{fig2}
\end{figure*}

In this work, we focus our analysis on SFGs. SFGs are selected as those with log sSFR$>{-11.0}~\rm yr^{-1}$, where sSFR=${\rm SFR}/M_{\ast}$. 3590 SFGs are finally selected. In Figure~\ref{fig2}, we show $\nabla_{\rm age}$ as a function of $M_{\ast}$ for the selected SFG sample. The mean value of $\nabla_{\rm age}$ is negative, which is around $-0.136~\rm log~Gyr{/R_{e}}$. This indicates that SFGs generally have older central stellar population than their outskirts, consistent with the "inside-out" galaxy formation scenario \citep{Wang 2011,Gonzalez 2015,Pan 2015}.

To verify our SFG selection, we cross-match the MaNGA sample with the GALEX$-$SDSS$-$WISE Legacy Catalog (GSWLC, see \citealt{Salim 2016}), which derive SFRs based on broad-band ultraviolet+infrared photometry. Of the 3590 MaNGA selected SFGs, 2579 ones have GSWLC counterparts. We find that 2471 ones ($\sim 95$\%) also have log $\rm {sSFR_{GSWLC}>{-11.0}~yr^{-1}}$. This indicates that the SFG sample selected by $\rm {sSFR_{H_{\alpha}}}$ is almost all overlapped with that selected by $\rm {sSFR_{\rm GSWLC}}$.

\begin{figure*}
\centering
\includegraphics[width=160mm,angle=0]{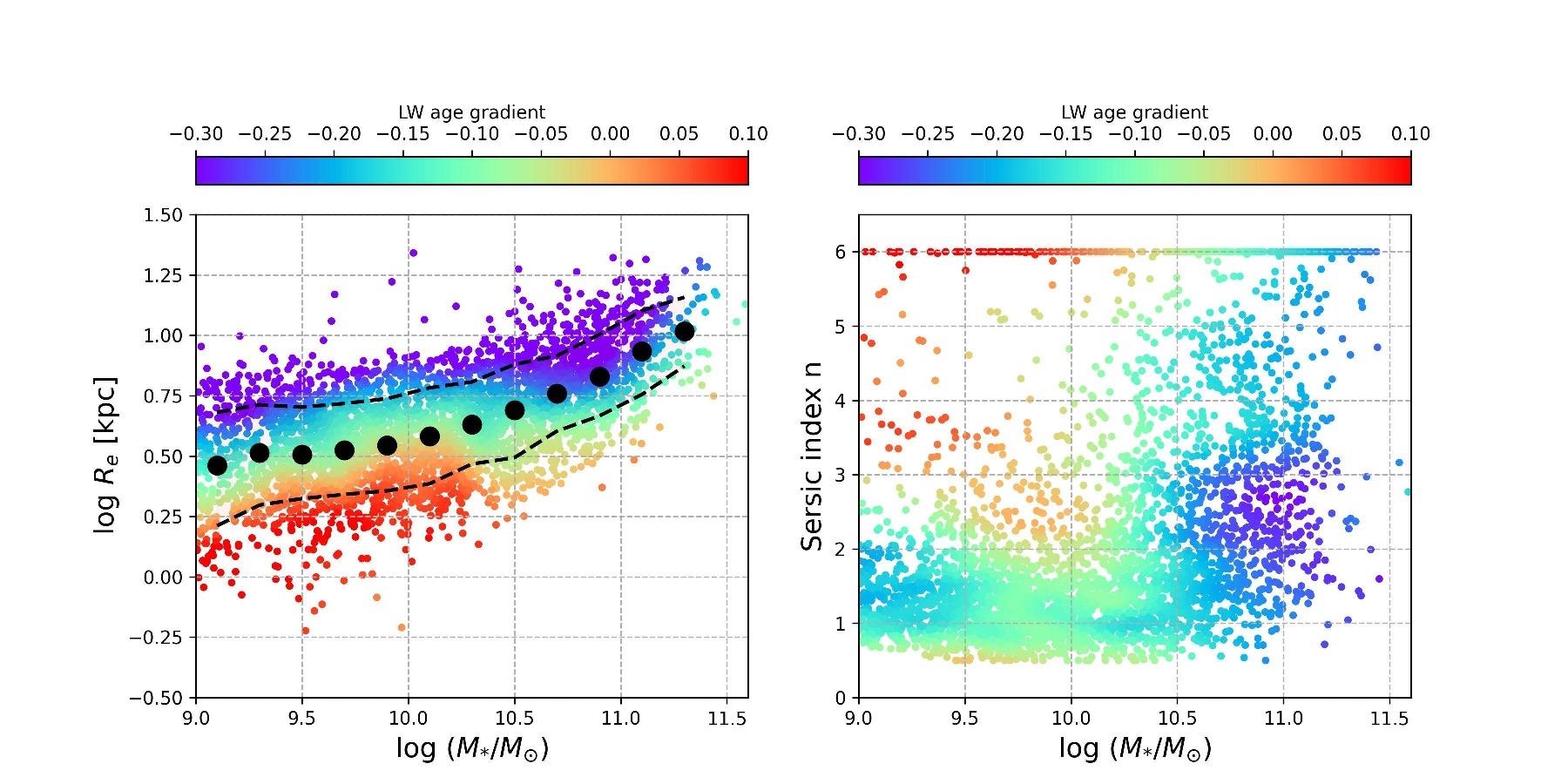}
\caption{Left:The $R_{\rm e}-M_{\ast}$ relation, color coded by the luminosity weighted stellar age gradient (with LOESS smoothing). Large symbols show the median $R_{\rm e}$ values of each stellar mass bin, with a bin size of $\Delta M_{\ast}=0.2$ dex. The two dashed lines indicate the $16-84$ percentile region. Right: S\'{e}rsic  index $n$ as a function of $M_{\ast}$, with LOESS smoothing. Symbols are color coded by luminosity weighted stellar age gradient. }\label{fig3}
\end{figure*}

We further divide the sample into stellar mass bins and investigate their $\nabla_{\rm age}$ distributions more specifically, with a bin size of $\Delta M_{\ast}=0.2~$dex. The result is shown in Figure~\ref{fig2}. It is clear that the $\nabla_{\rm age}$ distribution exhibits a "bump structure" near log$(M_{\ast}/M_{\sun})=10$. In this work, we consider SFGs with positive $\nabla_{\rm age}$ are undergoing/or have recently underwent central mass build-up processes, thus exhibiting younger central stellar ages than the outer regions. We then investigate the fraction of galaxies with positive $\nabla_{\rm age}$ (which we denote as $F_{\rm positive}$) in each stellar mass bin. To do this, the uncertainty of $\nabla_{\rm age}$ ($\sigma_{\nabla_{\rm age}}$) needs to take into account. We first construct the $\nabla_{\rm age}$ probability distribution function (PDF) for each SFG assuming a Gaussian form distribution centered at $\nabla_{\rm age}$, with a dispersion of $\sigma=\sigma_{\nabla_{\rm age}}$. With the constructed PDFs, we then perform bootstrapped resamplings to calculate $F_{\rm positive}$. We generate 500 realizations and derive the $F_{\rm positive}$ distribution at each stellar mass bin. The violin plot of the right panel of Figure~\ref{fig2} shows the result. As can be seen, the median $F_{\rm positive}$ is around 27\% at log$(M_{\ast}/M_{\sun})=9.1$, while it rises up to $\sim$ 40\% at log$(M_{\ast}/M_{\sun})\sim10$ and then rapidly decreases, reaching its minimum value of $\sim15\%$ near log$(M_{\ast}/M_{\sun})=11.0$. The peculiar distribution at log$(M_{\ast}/M_{\sun})=11.3$ may be of less statistical significance due to the small number of galaxies in this mass bin. The violin plot suggests that the excess of $F_{\rm positive}$ at log$(M_{\ast}/M_{\sun})\sim 10$ is of high significance.

The low fraction of SFGs with positive age gradients exhibits at log$(M_{\ast}/M_{\sun})>10.5$ is expected, since massive SFGs typically contain old bulge components \citep{Pan 2016,Belfiore 2017}. An interesting feature revealed in Figure~\ref{fig2} is that, low-mass SFGs seem to have a relatively low central mass build-up efficiency compared to those near log$(M_{\ast}/M_{\sun})=10$. We will come back to this issue in the discussion section.

\begin{figure*}
\centering
\includegraphics[width=160mm,angle=0]{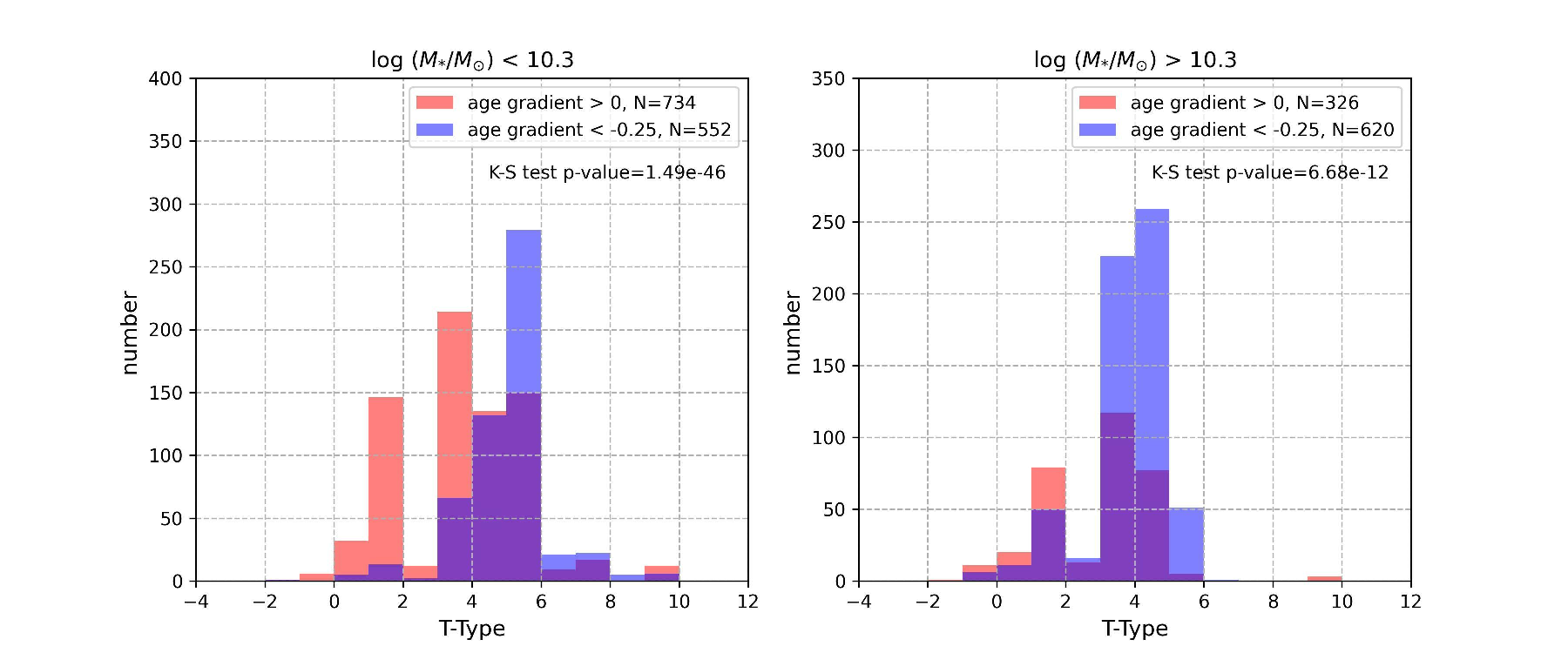}
\caption{The T-Type distribution for $\nabla_{\rm age}>0$ and $\nabla_{\rm age}<-0.25$ subsamples. We show the results in two stellar mass bins, with log$(M_{\ast}/M_{\sun})<10.3$ (left panel) and log$(M_{\ast}/M_{\sun})>10.3$ (right panel). The p-value of Kolmogorov-Smirnov test is also marked.}\label{fig4}
\end{figure*}

\subsection{The dependence of luminosity-weighted age gradient distribution on galaxy morphology }
Previous works reported that the radial gradients of many galactic properties are dependent upon galaxy morphologies \citep{Gonzalez 2015,Li 2018,Parikh 2021}. In this section, we investigate the dependence of $\nabla_{\rm age}$ on the morphologies of SFGs. In the left panel of Figure~\ref{fig3}, we explore the distribution of $\nabla_{\rm age}$ across the $R_{\rm e}-M_{\ast}$ plane. At fixed $M_{\ast}$, SFGs with small $R_{\rm e}$ tend to have shallow or positive $\nabla_{\rm age}$. A similar trend is also exhibited across the $\Sigma_{*,1kpc}-M_{\ast}$ plane of SFGs \citep{Woo 2019}, where $\Sigma_{*,1kpc}$ is the stellar mass density in the central 1kpc region of a galaxy. The large symbols show the median sizes at fixed $M_{\ast}$, i.e, the size$-$mass relation. As can be seen, the slope of the size$-$mass relation is shallow at log$(M_{\ast}/M_{\sun})<10.0$ and becomes steepened and tighter toward the high mass end. In the right panel, we investigate the $\nabla_{\rm age}$ distribution across the $n-M_{\ast}$ plane. Data points with $n=6$ correspond to the largest value allowed in the S\'{e}rsic profile fitting \citep{Blanton 2005}. Overall, SFGs with large $n$ tend to have shallow or positive $\nabla_{\rm age}$. At fixed $M_{\ast}$, $\nabla_{\rm age}$ is more strongly correlated with $R_{\rm e}$ than $n$. This may be due to the fact that a single S\'{e}rsic function is not a good parameterization of galaxy morphology \citep{Tasca 2011}. 

In Figure~\ref{fig4}, we investigate the T-Type distributions for SFGs within two $\nabla_{\rm age}$ bins, $\nabla_{\rm age}>0$ and $\nabla_{\rm age}<-0.25$. As can be seen, SFGs with $\nabla_{\rm age}>0$ tend to have a broader T-Type distribution than those with $\nabla_{\rm age}<-0.25$. Also, SFGs with $\nabla_{\rm age}>0$ contain more fractions of galaxies with T-Type $\leq$ 3. This trend is particularly evident at log$(M_{\ast}/M_{\sun})<10.3$, which is consistent with the trend shown in the left panel of Figure~\ref{fig3}. We perform a Kolmogorov-Smirnov (K-S) test to the T-Type distributions of the two samples. The small p-values of the K-S test suggest that these two distributions are statistically different. To conclude, at fixed $M_{\ast}$, SFGs with positive $\nabla_{\rm age}$ tend to have more centrally concentrated morphologies, consistent with the central mass build-up picture.

\begin{figure*}
\centering
\includegraphics[width=160mm,angle=0]{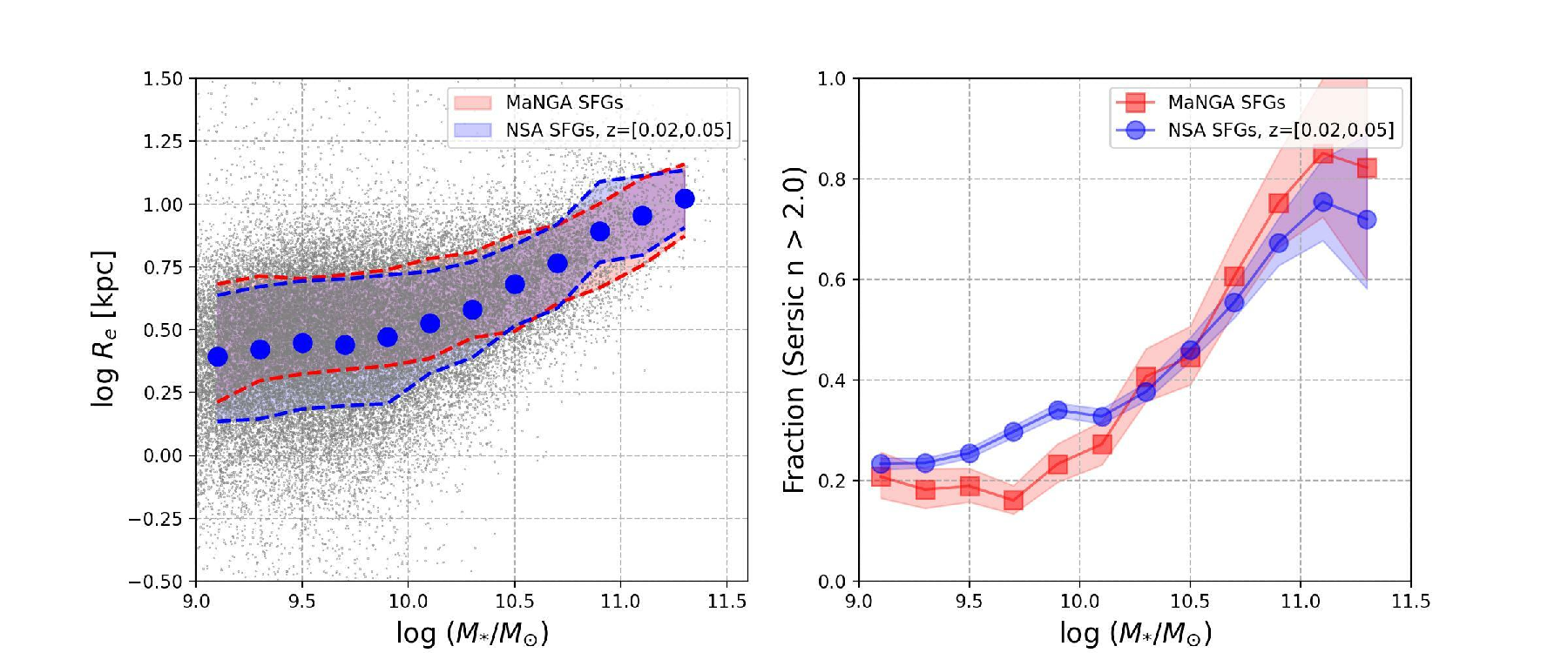}
\caption{Left: The $R_{\rm e}-M_{\ast}$ relation for the NSA SFG sample selected at $0.02<z<0.05$. Large symbols show the median $R_{\rm e}$ values of each stellar mass bin, with a bin size of $\Delta M_{\ast}=0.2$ dex. The two blue dashed lines indicate the $16-84$ percentile region. The red region indicates the $16-84$ region of MaNGA SFGs as shown in Figure~\ref{fig3}. Right: the fraction of SFGs with $n>2$ in the MaNGA and NSA-sloan sample. The shaded region indicates the $1\sigma$ region.}\label{fig5}
\end{figure*}

\begin{figure*}
\centering
\includegraphics[width=160mm,angle=0]{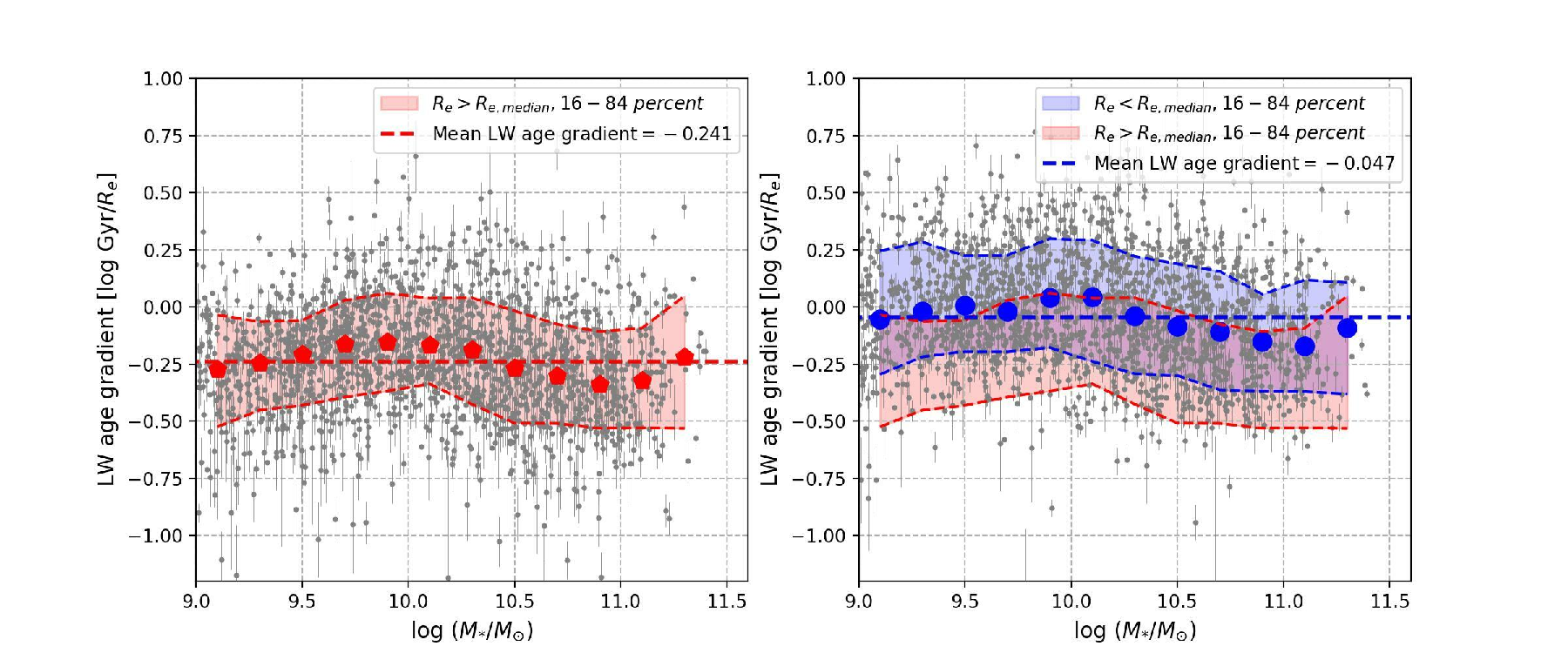}
\caption{Left: The luminosity weighted age gradient as a function of $M_{\ast}$ for SFGs with $R_{\rm e}>R_{\rm e, median}$. The mean age gradient is marked by the red dashed line. Large symbols show the median $\nabla_{\rm age}$ values of each stellar mass bin, with a bin size of $\Delta M_{\ast}=0.2~$dex. The shaded region indicates the region of $16-84$ percentile. Right: similar to the left panel, but for SFGs with $R_{\rm e}<R_{\rm e, median}$. The red region indicates the region of $16-84$ percentile for the $R_{\rm e}>R_{\rm e, median}$ subsample. }\label{fig6}
\end{figure*}

\section{Discussion}
\subsection{Is the MaNGA SFG sample representative?}

One may ask whether the MaNGA SFG sample is representative for the local SFGs in terms morphologies. To answer this question, we compare the $R_{\rm e}-M_{\ast}$ relation and the $f_{\rm n>2}-M_{\ast}$ relation of MaNGA SFGs with that of a local SFG comparison sample, where $f_{\rm n>2}$ is the fraction of SFGs with S\'{e}rsic index $n>2$. In the literature, galaxies with $n>2.5$ are considered to harbor prominent bulges, while those with $n<1.5$ are considered as bulgeless galaxies \citep{Blanton 2003,Bell 2012}. Inspecting the $n-M_{\ast}$ relation, we consider galaxies with $n>2$ have built a bulge component. The comparison SFG sample is selected from the NSA-Sloan catalog, which is the master photometric catalog of MaNGA Pipe3D. We first cross-match the NSA catalog with the GSWLC catalog to drive SFR measurements. Then we select SFGs with $b/a>0.4$ at $0.02<z<0.05$ as the comparison sample, with $\rm log~sSFR_{\rm GSWLC}>-11.0~yr^{-1}$. In this redshift range, the SDSS spectroscopic main galaxy sample is completed to $M_{*}\sim 10^{9}M_{\sun}$ \citep{Schawinski 2010}. As mentioned before, the $\rm sSFR_{\rm GSWLC}$ selected SFGs are almost all overlapped with those selected by $\rm sSFR_{\rm H\alpha}$, thus this comparison should be physically meaningful. The comparison SFG sample contains $\sim 37000$ galaxies.

We show the result in Figure~\ref{fig5}. As can be seen, the $R_{\rm e}-M_{\ast}$ relation of MaNGA SFGs is consistent with that of the NSA catalog at log$(M_{\ast}/M_{\sun})>10.0$. At log$(M_{\ast}/M_{\sun})<10.0$, MaNGA SFGs are biased against compact galaxies. On the one hand, this is partially due to the fact that we have removed the 19-fiber IFUs in the sample selection. We confirmed that when the 19-fiber IFUs are included, the discrepancy between the two samples in the $R_{\rm e}-M_{\ast}$ relation is reduced at log$(M_{\ast}/M_{\sun})<10.0$. On the other hand, this should be mainly due to the selection effect of MaNGA, because low-mass compact galaxies are difficult to spatially resolve in the MaNGA observation. Such a trend is confirmed by the $f_{\rm n>2}-M_{\ast}$ relation, where the $f_{\rm n>2}$ of MaNGA SFGs is lower than that of the NSA sample at log$(M_{\ast}/M_{\sun})<10.0$.

Figure~\ref{fig5} suggests that at fixed $M_{\ast}$, galaxies located in the high $R_{\rm e}$ region should be less affected by the selection effect of MaNGA. We thus divide the MaNGA SFG sample into two subsamples according to their $R_{\rm e}$, one with $R_{\rm e}>R_{\rm e,~median}$ and the other with $R_{\rm e}<R_{\rm e,~median}$. We show the $\nabla_{\rm age}-M_{\ast}$ relation for these two samples in Figure~\ref{fig6}. As can be seen, the bump structure at log$(M_{\ast}/M_{\sun})\sim10.0$ exhibits in both subsamples. Interestingly, it seems that the bump is more prominent in the $R_{\rm e}>R_{\rm e,~median}$ subsample, for which the result should be less affected by the MaNGA selection effect. To conclude, although the MaNGA SFG sample is biased against low-mass compact galaxies, our findings should not be significantly affected by this sample selection effect.

\begin{figure*}
\centering
\includegraphics[width=160mm,angle=0]{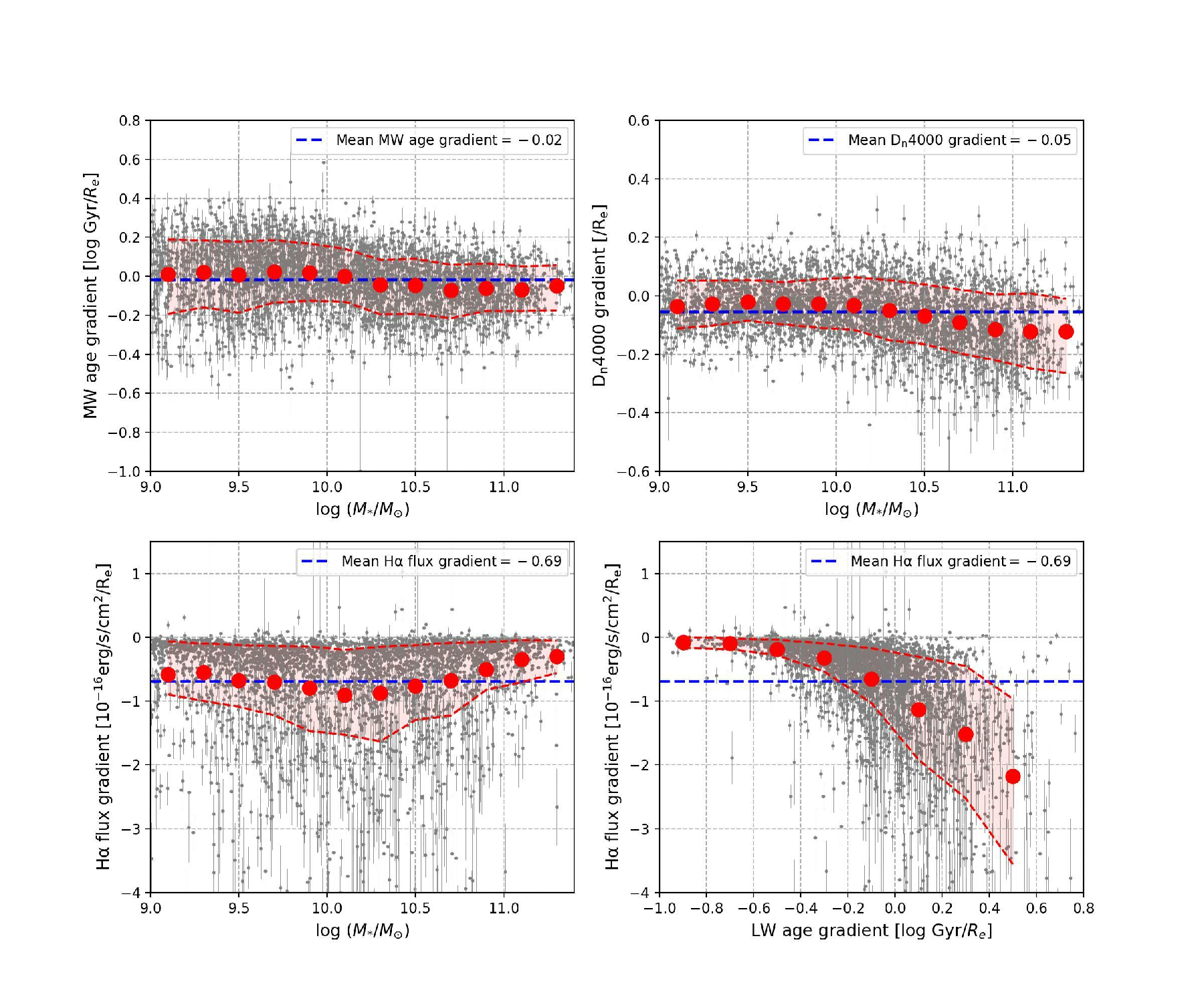}
\caption{Top left: the mass-weighted age gradient as a function of $M_{\ast}$. The stellar mass bin size is $\Delta M_{\ast}=0.2$ dex. Large symbols show the median values of each stellar mass bin. Top right: D4000 gradient as a function of $M_{\ast}$. Bottom left: the H$\alpha$ flux gradient as a function of $M_{\ast}$. Bottom right: the H$\alpha$ flux gradient as a function of luminosity-weighted age gradient. large symbols show the median values of each $\nabla_{\rm age}$ bin. In each panel, the shaded region indicates the $16-84$ percentile region.}\label{fig7}
\end{figure*}

\subsection{Using other stellar age indicators}
In this section, we first check the robustness of our analysis that based on luminosity-weighted age gradient. In the top left panel of Figure~\ref{fig7}, we show the mass-weighted age gradient as a function of $M_{\ast}$. As can be seen, the mass-weighted age gradient exhibits no "bump structure" near log$(M_{\ast}/M_{\sun})=10$, in stark contrast to Figure~\ref{fig2}. The mean mass-weighted age gradient is also much shallower compared to the mean luminosity-weighted age gradient. This is because the mass-weighted quantities typically have narrower dynamical range than the luminosity-weighted quantities. In the top right panel, we show the 4000 \AA~break $D_{n}(4000)$ gradient as a function of $M_{\ast}$. $D_{n}(4000)$ is a commonly used stellar age indicator, which is sensitive to star formation activity on timescales of $\sim 1-2$ Gyr \citep{Kauffmann 2003, Wang 2018}. The $D_{n}(4000)$ gradient is flat at log$(M_{\ast}/M_{\sun})<10.2$ and turns negative toward the massive end, consistent with previous findings \citep{Wang 2018}. Overall, the behavior of $D_{n}(4000)$ gradient follows that of the mass-weighted stellar age gradient, showing no excess of central mass build-up events near $M_{\ast}=10^{10}M_{\sun}$.

In the bottom left panel of Figure~\ref{fig7}, we show the gradient of H$\alpha$ flux as a function of $M_{\ast}$. H$\alpha$ emission is sensitive to star formation activity on timescales of several Myr and is a traditional tracer of SFR. Thus, a very steep H$\alpha$ flux gradient indicates a high star formation concentration. Interestingly, the H$\alpha$ flux gradient steepens near log$(M_{\ast}/M_{\sun})=10.0$, which echoes the "bump structure" we identify in Figure~\ref{fig2}. In the bottom right panel of Figure~\ref{fig7}, we show that SFGs with positive $\nabla_{\rm age}$ indeed have steepened H$\alpha$ flux gradient, indicative of more centrally concentrated star formation. To conclude, Figure~\ref{fig7} suggests that signs of central mass build-up events can be better detected using the radial gradients of star formation tracers on short timescales. In stead, mass-weighted stellar age and $D_{n}(4000)$ gradients are less sensitive to such events, possibly due to their narrow dynamical range.

\subsection{Bulge formation in different mass regime}
At log$(M_{\ast}/M_{\sun})<9.5$, $\sim80$\% SFGs have $n<2$, i.e., the majority of low-mass SFGs have not built a central bulge component. In observation, low-mass SFGs are rich in neutral hydrogen (\Hi) gas \citep{Saintonge 2022}. As shown in previous works, \Hi~gas plays an important role in the build-up of a disk component \citep{Wang 2011, Chen 2020, Pan 2021}.  With a high \Hi~gas fraction, disk rebuilding could be quit common for low-mass compact SFGs. In addition, major merger events, which are considered to be an important channel for bulge formation, are rare in this mass regime \citep{Casteels 2014,Rodriguez 2015}. Finally, the bulge formation efficiency of low-mass galaxies can be suppressed by their high gas richness during merger events \citep{Hopkins 2009,Hopkins 2010}. These effects all potentially contribute to a low central mass build-up efficiency as seen in the low-mass regime.

At log$(M_{\ast}/M_{\sun})>10.5$, the majority of SFGs have built a bulge component (with $n>2$). For massive galaxies, major mergers are considered to play an important role in bulge formation \citep{Hopkins 2010}. Recent studies showed that stellar migration may also contribute significantly to the build-up of central mass for Milky Way-mass galaxies \citep{Boecker 2023}. The relative contribution of different mechanisms to the bulge formation of massive galaxies is a matter of intense debate. Regardless the detail formation mechanism, many bulges of massive SFGs show very low star formation activity, indicating that the central mass build-up process is largely completed, and the galaxies are on the path of star formation quenching \citep{Belfiore 2017}.

SFGs with intermediate mass stand in a unique position in bulge formation. In the mass range of $9.5<{\rm log}(M_{\ast}/M_{\sun})<10.5$, neither "mass quenching"  is operated effectively \citep{Peng 2010}, nor the SFGs are too rich in \Hi~gas. We indeed find evidence that SFGs near log$(M_{\ast}/M_{\sun})=10.0$ build up their central mass concentration most actively. Such a phenomenon is helpful to interpret the systematic morphological differences between low-mass and high-mass SFGs, as shown in Figure~\ref{fig5}. We also note that our results are based on the low-redshift SFG sample. The bulge formation picture could be different at high redshift, as high-z galaxies exhibit more intense and clumpy star formation. In future work, it would be interesting to extend our study to higher redshifts to explore bulge formation across cosmic time.

\section{Conclusion}
In this paper, we investigate the luminosity-weighted age gradient distribution of $\sim3600$ local SFGs based on the MaNGA Pipe3D dataset. The mean age gradient is negative, with $\nabla_{\rm age}=-0.14$~log Gyr/$R_{e}$, consistent with the inside-out disk formation scenario. Specifically, SFGs with positive $\nabla_{\rm age}$ consist of $\sim 28\%$ at log$(M_{\ast}/M_{\sun})<9.5$, while this fraction rises up to its peak ($\sim 40\%$) near log$(M_{\ast}/M_{\sun})=10$ and then rapidly decreases, reaching its minimum value of $\sim15\%$ near log$(M_{\ast}/M_{\sun})=11.0$. We find that galaxies with positive $\nabla_{\rm age}$ typically have more compact sizes and more centrally concentrated star formation than their counterparts, indicating that they are undergoing/or recently underwent the compaction process to build up central mass concentration. We conclude that the build-up of central stellar mass concentration in local SFGs is mostly active near $M_{\ast}=10^{10}M_{\sun}$. These findings are helpful to interpret the systematic morphological difference across the SFG population from low-mass to high-mass regime.

\acknowledgments
%\begin{acknowledgments}
We are grateful to the anonymous referee for very useful comments that have improved the clarity of this paper. This work was partially supported by  the National Natural Science Foundation of China (NSFC, Nos. 12173088, 12233005 and 12073078), the science research grants from the China Manned Space Project with NO. CMS-CSST-2021-A02, CMS-CSST-2021-A04 and CMS-CSST-2021-A07.

Funding for the Sloan Digital Sky Survey IV has been provided by the Alfred P. Sloan Foundation, the U.S. Department of Energy Office of Science, and the Participating Institutions.SDSS-IV acknowledges support and resources from the Center for High Performance Computing  at the University of Utah. The SDSS website is www.sdss4.org. SDSS-IV is managed by the Astrophysical Research Consortium for the Participating Institutions of the SDSS Collaboration including the Brazilian Participation Group, the Carnegie Institution for Science, Carnegie Mellon University, Center for Astrophysics | Harvard \& Smithsonian, the Chilean Participation Group, the French Participation Group, Instituto de Astrof\'isica de Canarias, The Johns Hopkins University, Kavli Institute for the Physics and Mathematics of the Universe (IPMU) / University of Tokyo, the Korean Participation Group, Lawrence Berkeley National Laboratory, Leibniz Institut f\"ur Astrophysik Potsdam (AIP),  Max-Planck-Institut f\"ur Astronomie (MPIA Heidelberg), Max-Planck-Institut f\"ur Astrophysik (MPA Garching), Max-Planck-Institut f\"ur Extraterrestrische Physik (MPE), National Astronomical Observatories of China, New Mexico State University, New York University, University of Notre Dame, Observat\'ario Nacional / MCTI, The Ohio State University, Pennsylvania State University, Shanghai Astronomical Observatory, United Kingdom Participation Group,Universidad Nacional Aut\'onoma de M\'exico, University of Arizona, University of Colorado Boulder, University of Oxford, University of Portsmouth, University of Utah, University of Virginia, University of Washington, University of Wisconsin, Vanderbilt University, and Yale University.

%\end{acknowledgments}

\software{astropy \citep{astropy 2013,astropy 2018}, matplotlib  \citep{Hunter 2007} }

\begin{figure*}
\centering
\includegraphics[width=160mm,angle=0]{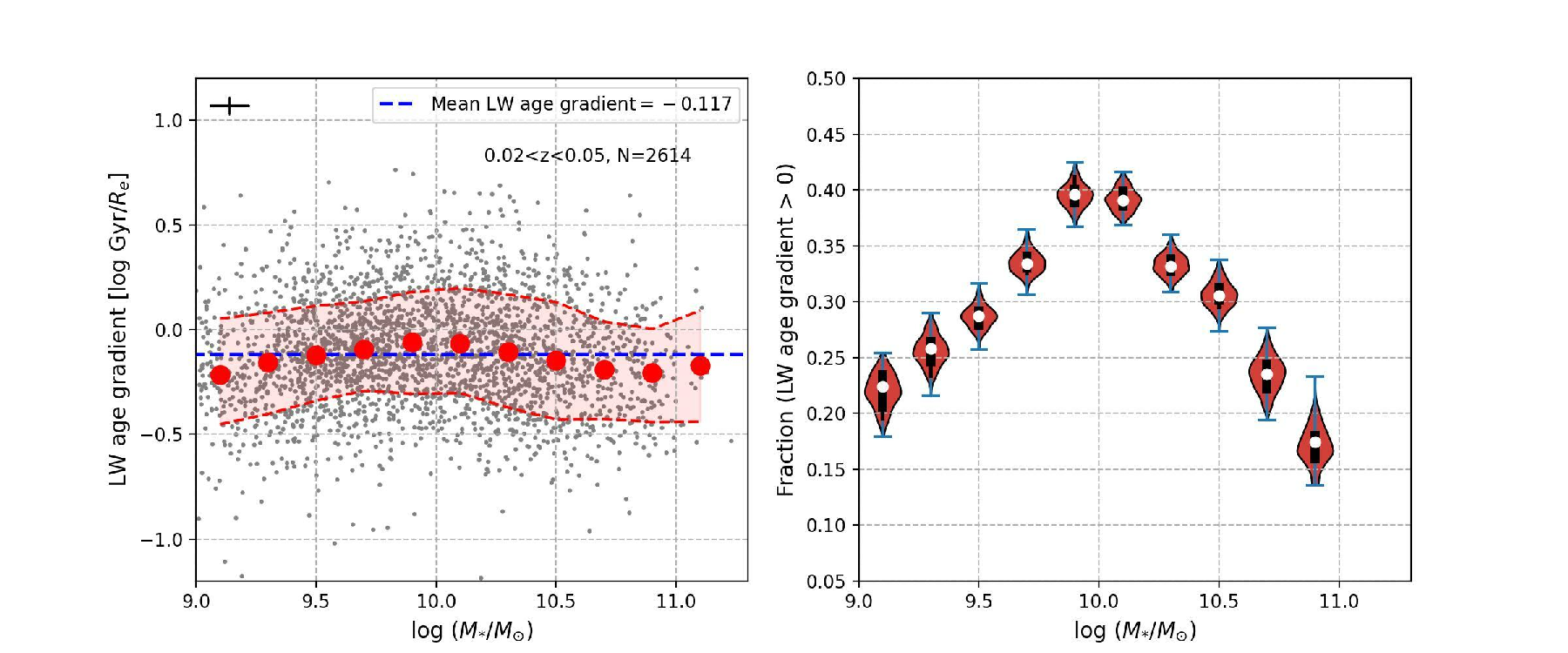}
\caption{Similar to Figure 2, with the SFG sample of $0.02<z<0.05$.}\label{fig8}
\end{figure*}

\appendix\label{sec:app}
In the main body of this work, we select galaxy sample with a redshift range of $0.01<z<0.15$, which corresponds to a cosmic time interval of $\sim1.8$ Gyr. To investigate whether our results are biased due to the redshift range adopted, we select SFGs within a narrower redshift range ($0.02<z<0.05$) and repeat our analysis. In Figure~\ref{fig8}, we show the age gradient as a function of stellar mass with $\sim 2600$ SFGs at $0.02<z<0.05$. As can be seen, the main feature is unchanged compared to Figure~\ref{fig2}. Thus the results presented in the main body of the article should not be affected by the adopted redshift range in the sample selection.

\end{document}